\documentclass[11pt]{JHEP3}
\usepackage{amsmath,amssymb,epsfig}
\usepackage{amsfonts}
\usepackage{verbatim}
\usepackage{latexsym}
\usepackage{amsthm}
\usepackage{amsbsy}
\usepackage{xspace}

\def\one{{\,\hbox{1\kern-.8mm l}}}
\newcommand{\Dslash}{\not{\hbox{\kern-4pt $D$}}}
\newcommand{\pdslash}{\not{\hbox{\kern-2pt $\partial$}}}
 
 \newcommand{\cO}{\mathcal{O}}

\newcommand{\Comment}[1]{{}}

\def\IZ{{\mathbb Z}}

\def\IR{{\mathbb R}}

\def\calo         {{\cal O}}

\def\calv         {{\cal V}}

\newcommand{\bc}{\begin{center}}
\newcommand{\ec}{\end{center}}
\newcommand{\ba}{\begin{array}}
\newcommand{\ea}{\end{array}}
\newcommand{\beq}{\begin{equation}}
\newcommand{\eeq}{\end{equation}}
\newcommand{\bea}{\begin{eqnarray}}
\newcommand{\eea}{\end{eqnarray}}
\newcommand{\bmx}{\begin{pmatrix}}
\newcommand{\emx}{\end{pmatrix}}

\newcommand{\be}{\begin{equation}}
\newcommand{\ee}{\end{equation}}

\newcommand{\del}{\partial}
\newcommand{\half}{{\frac{1}{2}\,}}
\newcommand{\tr}{{\rm tr}}

\newcommand{\eref}[1]{Eq.\,(\ref{#1})}

\newcommand{\zbar}{{\bar z}}
\newcommand{\wbar}{{\bar w}}
\newcommand{\vbar}{{\bar v}}
\newcommand{\qbar}{{\bar q}}

\newcommand{\Dbar}{{\bar D}}
\newcommand{\calvbar}{{\bar {\cal V}}}
\newcommand{\phibar}{{\bar \phi}}

\newcommand{\alphabar}{{\bar \alpha}}
\newcommand{\Deltabar}{{\bar \Delta}}
\newcommand{\cT}{{\cal T}}

\newcommand{\valpha}{{\vec\alpha}}

\newcommand{\vlambda}{{\vec\lambda}}

\newcommand{\tS}{{\tilde S}}
\newcommand{\tZ}{{\tilde Z}}

\newcommand{\llangle}{\langle\!\langle}
\newcommand{\rrangle}{\rangle\!\rangle}
\newcommand{\Renyi}{R\'enyi\xspace}
\def\IB{\relax{\rm I\kern-.18em B}}
\def\IC{{\relax\hbox{\kern.3em{\cmss I}$\kern-.4em{\rm C}$}}}
\def\ID{\relax{\rm I\kern-.18em D}}
\def\IE{\relax{\rm I\kern-.18em E}}
\def\IF{\relax{\rm I\kern-.18em F}}
\def\II{\relax{\rm I\kern-.18em I}}
\def\IZ{\relax{\sf Z\kern-.35em Z}}
\def\Id{\relax{1\kern-.32em 1}}
\def\IG{\relax\hbox{$\inbar\kern-.3em{\rm G}$}}
\def\IR{\relax{\rm I\kern-.18em R}}
\newcommand\sfrac[2]{{\textstyle\frac{#1}{#2}}}

\newcommand\shalf{{\textstyle\frac12}}

\normalsize

\title{Modular invariance and entanglement entropy}

\author{
Sagar Fakirchand Lokhande\footnote{email: sagar.f.lokhande@gmail.com, sunil.mukhi@gmail.com}~ and Sunil Mukhi$^*$\footnote{On leave from the Tata Institute of Fundamental Research, Mumbai.}\\~\\
{\it Indian Institute of Science Education and Research,\\ 
Homi Bhabha Rd, Pashan, Pune 411 008, India}
}

\abstract{We study the \Renyi and entanglement entropies for free 2d CFT's at finite temperature and finite size, with emphasis on their properties under modular transformations of the torus. We address the issue of summing over fermion spin structures in the replica trick, and show that the relation between entanglement and thermal entropy determines two different ways to perform this sum in the limits of small and large interval. Both answers are modular covariant, rather than invariant. Our results are compared with those for a free boson at unit radius in the two limits and complete agreement is found, supporting the view that entanglement respects Bose-Fermi duality. We extend our computations to multiple free Dirac fermions having correlated spin structures, dual to free bosons on the Spin(2d) weight lattice. }

\preprint{}


\begin{document}

\section{Introduction}

Entanglement entropy\cite{Holzhey:1994we,Casini:2004bw,Calabrese:2004eu} has become a powerful diagnostic tool for a variety of phenomena in quantum theory (see for example Refs.\cite{2002Natur.416..608O,Klebanov:2007ws,Riera:2006vj}). Very important progress in our understanding of entanglement arose from the proposal \cite{Ryu:2006bv} of holographic entanglement entropy, which has beautifully linked the study of entanglement to gravity and thermodynamics (for example, the recent work of Ref.\cite{Faulkner:2013ica}).

To define entanglement we break up the degrees of freedom into two disjoint sets $A$ and $B$. The density matrix of the full system is taken to be $\rho$. Define the reduced density matrix in the region $A$ as $\rho_A=\tr_B\rho$, obtained by tracing out the degrees of freedom in region $B$. Then the entanglement entropy is defined as:
\be
S_E=-\tr\rho_A\log\rho_A
\ee
A more general quantity called the \Renyi entropy is often studied. This is defined as:
\be
S_n = \frac{1}{1-n}\log\tr \rho_A^n
\ee
where $n$ is a positive integer. Besides being a diagnostic on its own, the \Renyi entropy can sometimes be used as a trick to compute the entanglement entropy. This happens when $S_n$ can be unambiguously continued to arbitrary real values of $n$. Then it is easily verified that the limit $\lim_{n\to 1}S_n = S_{E}$. The advantage of focusing on \Renyi entropy is that it can often be computed using the replica trick\cite{Casini:2005rm, Cardy:2007mb,Casini:2009sr,Calabrese:2009qy}. For this,  one constructs ``twist fields'' and computes their correlation functions using relatively standard procedures in CFT. The twist correlators determine a quantity called the ``replica partition function'' $Z_n$, in terms of which the \Renyi entropy is:
\be
S_n =\frac{1}{1-n}\log \frac{Z_n}{Z_1^n}
\ee

Entanglement and \Renyi entropies have been computed for a wide variety of quantum field theories but they tend to have the problem that either (i) they are very universal, depending on limited information about the QFT (in which case they are not a very useful diagostic), or (ii) they are extremely hard to compute. In class (i) one has conformal field theories at finite temperature {\em or} in a finite space, with a single entangling interval, for which the entanglement entropy depends only on the central charge. However when the temperature and spatial size are both finite, there is a finite space with multiple entangling intervals, the same theories do not exhibit such a level of universality and therefore this situation is more interesting. The case of multiple intervals at zero temperature was investigated in Refs.\cite{Casini:2008wt,Calabrese:2009ez,Calabrese:2010he,Headrick:2012fk}. At finite temperature and finite size (i.e., on a Euclidean torus), only a few explicit computations exist and these apply to free field theories of bosons or fermions. The  results for the \Renyi entropy are found to depend sensitively on the spectrum of operator dimensions and not just the central charge. To date, explicit computations have been carried out only for the free Dirac fermion\cite{Azeyanagi:2007bj,Herzog:2013py} and a pair of free scalar fields \cite{Datta:2013hba,Chen:2014hta,Chen:2015cna}. There are also some general results about the universal thermal correction in generic 2d CFT's\cite{Cardy:2014jwa,Chen:2014unl}. Quantum corrections to such entropies have been addressed in the holographic context in Refs.\cite{Barrella:2013wja,Faulkner:2013ana}.

Once we are on the torus (defined by the inverse temperature $\beta$ and the system size $L$), the issue of modular invariance naturally arises. Early computations including those in\cite{Azeyanagi:2007bj,Herzog:2013py}, obtained the 
\Renyi entropy for free Dirac fermions in a fixed spin structure (which could be any of $(-,-), (+,-), (-,+)$ but not $(+,+)$). The computation made use of local Bose-Fermi equivalence to construct the twist field and compute its correlators, then manually restricted the bosonic sum to obtain a fixed fermion spin structure. The results are not modular-invariant and therefore do not respect Bose-Fermi duality. This brings into question whether \Renyi and entanglement entropies are properties of a QFT independent of its presentation, or depend on the specific fields used to define the QFT. 

This point is discussed in Ref.\cite{Headrick:2012fk}\footnote{We thank Shiraz Minwalla for bringing this work to our attention.} which makes a strong case that entanglement should depend only on the theory rather than on its presentation. The consistent definition of a CFT incorporates modular invariance as a fundamental property and this  provides a valuable constraint on its operator content, partition function and correlators \cite{Seiberg:1986by, Cardy:1991kr}. Motivated by this viewpoint we revisit the previous computation of \Renyi entropy for free fermions, performed using the replica trick, and examine whether a modular-invariant replica partition function $Z_n$ exists satisfying all the desired properties. In order to achieve modular invariance one needs to perform a sum over spin structures (i.e. over anti-periodic and periodic boundary conditions in both the spatial and thermal directions) \cite{Seiberg:1986by}.  

The replica trick involves extending the original theory to an $n$-fold copy and performing a twist-field computation on this copy. The presence of fermion spin structures introduces, quite literally, a new twist on the problem -- one has to decide whether the spin structures are correlated or uncorrelated across replicas. To get some physical input into which one is correct, we appeal to the universal relation between entanglement and thermal entropy first noted in Ref.\cite{Azeyanagi:2007bj}. We find a perhaps surprising result: for free fermions, this relation holds only if we choose two different ways of summing in two different limits: for a small entangling region one must consider uncorrelated spin structures, while for a large one (close to the size of the entire space) one has to completely correlate spin structures across replicas. The correct formula for a general size of the entangling region must interpolate between these two extremes, and we speculate at the end about how this could come about.

It is known that for large intervals, the replica trick requires a correlation between states on different replicas, as was originally evident from the work of \cite{Cardy:2014jwa}. In a beautiful series of works, this relation has been developed further and a formal proof been provided for the entanglement-thermal entropy relation\cite{Chen:2014ehg,Chen:2014hta}. Our computation can be seen as a verification of these ideas in the specific context of fermion spin structures. In Ref.\cite{Chen:2015cna} the proof has been explicitly verified for the free-boson partition function. We confirm that our result agrees with that of the above reference in the special case of radius $R=1$ (in units where $\alpha'=2$), where the free boson is known to be equivalent to a free Dirac fermion. This provides strong evidence, for the first time, that Bose-Fermi duality is respected by entanglement entropy on the torus.

Both the ways of summing over spin structures that we use lead to a replica partition function that is modular covariant rather than invariant. Under modular transformations of the torus, it transforms into itself with a prefactor. As a result the \Renyi and entanglement entropies acquire an additive contribution upon modular transformations. We will see that 
the covariance arises from the fact that twist field correlators are involved: correlation functions in CFT are modular covariant rather than invariant, and the precise prefactor that we find agrees with the general expectation. We verify that the \Renyi entropy for a compact free boson at radius $R$, originally attempted in \cite{Datta:2013hba} and recently corrected in Ref.\cite{Chen:2015cna}, is also modular covariant with the expected prefactor.

Next we turn our attention to more general free 2d CFT's. The CFT of two Dirac fermions with correlated spin structures is equivalent to a pair of bosons compactified at radius $R=\sqrt 2$ (this is the self-dual value under T-duality). We compute the \Renyi entropies for this theory for small and large intervals and find modular-covariant answers that satisfy the thermal entropy relation. This is then repeated for theories of multiple correlated Dirac fermions. These are CFT's of arbitrarily large central charge $d$ that are not direct sums of simpler CFT's. The bosonic duals are given by multiple free bosons compactified on the weight lattice of Spin(2d) with a specific background $B$-field\cite{Elitzur:1986ye}. We use the weight-lattice structure to construct a twist field, and compute its two-point function. Incorporating the sum over spin structures in the two relevant limits provides the replica partition function in these limits and confirms the arguments made above. We conclude with some observations and speculations about future directions.

\section{Free Dirac fermion and modular invariance}

Consider a CFT whose partition function on a rectangular torus of spatial size $L$ and Euclidean time extent $\beta$ is $Z_1(L,\beta)$. As mentioned above, the \Renyi entropy $S_n$ associated to a spatial region $A$ running from $0$ to $\ell$ is defined as:
\be
S_n = \frac{1}{1-n}\log\tr\,\rho_A^n
\label{renyidef}
\ee
where $\rho_A$ is the density matrix obtained by tracing out the degrees of freedom outside $A$. This can be evaluated using the replica trick, which we briefly summarise here. One extends the original torus to an $n$-fold cover with branch cuts along spatial intervals from $0$ to $\ell$. This can equivalently be thought of as a single copy of the original torus with the insertion of ``twist fields'' labelled $\sigma_k$  at the endpoints of the entangling interval. Here $k$ ranges from $-\frac{n-1}{2}$ to $\frac{n-1}{2}$ in integral steps. The job of the twist fields
is to introduce a phase factor $e^{\frac{2\pi i k}{n}}$ on the free fermion field, which we denote $D(z)$, as it goes around the twist field:
\be
\sigma_k(z,\zbar)D(w)\sim (z-w)^{\frac{k}{n}}
\label{twact}
\ee
and the opposite phase on the anti-holomorphic conjugate fermion:
\be
\sigma_k(z,\zbar)\Dbar(\wbar)\sim (\zbar-\wbar)^{-\frac{k}{n}}
\ee
The twist fields can be shown to satisfy $\sum_{k}\Delta_k = \frac{c}{24}\left(n-\frac{1}{n}\right)$ in any CFT, where $c$ is the central charge.

One can then show that\cite{Casini:2005rm,Calabrese:2009qy}
\be
\tr\,\rho_A^n = \prod_{k=-\frac{n-1}{2}}^{\frac{n-1}{2}}
\langle \sigma_k(\ell,\ell)\,\sigma_{-k}(0,0)\rangle
\ee
It is convenient to think of the product of {\em un-normalised} correlators as defining the ``replica partition function''
\be
Z_n = \prod_{k=-\frac{n-1}{2}}^{\frac{n-1}{2}}
Z_1 \langle \sigma_k(\ell,\ell)\,\sigma_{-k}(0,0)\rangle=
\prod_{k=-\frac{n-1}{2}}^{\frac{n-1}{2}}\llangle \sigma_k(\ell,\ell)\,\sigma_{-k}(0,0)\rrangle
\ee
where $Z_1$ is the ordinary partition function of the theory (without insertions), and we use $\llangle\cdots \rrangle$ to denote un-normalised correlators. It follows that:
\be
\tr\,\rho^n_A=\frac{Z_n}{Z_1^n}
\label{trrhon}
\ee
from which the \Renyi entropies are easily obtained.

Consider a Dirac fermion with a single entangling interval of length $\ell$. This theory consists of two Majorana fermions with correlated spin structures. Labelling the holomorphic parts of the Majorana fermions as $\psi_1,\psi_2$ we make the Dirac fermion $D(z) = \psi_1(z) +i\psi_2(z)$. The local operators of dimension $(\half,\half)$ arising from this are: $D(z)\Dbar(\zbar)$, $D^\dagger(z)\Dbar(\zbar)$, $D(z)\Dbar^\dagger(\zbar)$, $D^\dagger(z)\Dbar^\dagger(\zbar)$. This theory has the modular-invariant partition function:
\be
Z_{\rm Dirac}=\half\sum_{\nu=2,3,4}\left|\frac{\theta_\nu(0|\tau)}{\eta(\tau)}\right|^2
\ee
For future use we define:
\be
Z_1[m]=\half\sum_{\nu=2,3,4}\left|\frac{\theta_\nu(0|\tau)}{\eta(\tau)}\right|^{2m}
\label{zedonem}
\ee
where the subscript 1 refers to the ordinary partition function (this will later be replaced by $n$ for the $n$th replica partition function). This is the ordinary partition function for $m$ Dirac fermions with correlated spin structures. In this notation the single Dirac fermion has ordinary partition function $Z_1[1]$.

For the replica partition function of the same theory, the following result\footnote{In their conventions, $L=1$ and $\ell$ is denoted by $L$.} was obtained in 
Ref.\cite{Azeyanagi:2007bj}:
\be
\langle \sigma_k(z,\zbar)\sigma_{-k}(0,0)\rangle=
\left|\frac{\theta_1'(0|\tau)}{\theta_1(\frac{\ell}{L}|\tau)}\right|^{4\Delta_k}\left|\frac{\theta_\nu(\frac{k\ell}{nL}|\tau)}{\theta_\nu(0|\tau)}\right|^2
\label{azeyares}
\ee
Here $\Delta_k=\frac{k^2}{2n^2}$, $\nu=2,3,4$ is a fixed spin-structure on the torus corresponding to boundary conditions $(+,-),(-,-),(-,+)$ respectively\footnote{The first entry labels the spatial boundary condition and the second, the boundary condition in Euclidean time.}, and $\tau=i\frac{\beta}{L}$. For the $(+,+)$ spin-structure the result would formally be the same but with $\nu=1$. Subsequently the above results were re-derived in Ref.\cite{Herzog:2013py}, where it was observed that the 2-point function in the $(+,+)$ spin structure is divergent due to the vanishing of $\theta_1(0|\tau)$. Importantly, the entanglement entropy following from the computations in Refs.\cite{Azeyanagi:2007bj,Herzog:2013py} satisfies the relation proposed in Ref.\cite{Azeyanagi:2007bj}, based on holography, relating its small and large-interval limits to the thermal entropy of the same system.

Despite meeting all the other physical requirements, the above result raises a puzzle. As emphasised in Ref.\cite{Headrick:2012fk}, entanglement entropy should be a feature of a definite quantum field theory and independent of the presentation of that theory. If so, in a theory where Bose-Fermi correspondence holds, one should obtain the same result in both bosonic and fermionic formulations. A free fermion with a single spin structure is not dual to the theory of free bosons, rather it is the modular-invariant free fermion partition function (and correlation functions) that can be compared with those of the boson theory. In this spirit, we propose that \Renyi entropy be computed by performing a sum over all four spin structures in the replica partition function $Z_n$, and dividing by a corresponding quantity in the absence of replicas. In principle there is more than one way to do this, and this point will be central to our discussion.

Let us review the calculation of \cite{Azeyanagi:2007bj} for the Dirac fermon theory. To compute the correlators of fermionic twist fields on the torus in this theory, these operators are first identified with (non-local) operators in a free boson theory using the fact that a Dirac fermion is equivalent to a free compact boson of radius $R=1$ (in our conventions $\alpha'=2$, so T-duality acts by $R\to \frac{2}{R}$ and $R=\sqrt2$ is the self-dual radius). At a general radius $R$, the free compact boson has vertex operators labelled by integers $(e,m)$:
\be
{\cal O}_{e,m}(z,\zbar)=\calv_{e,m}(z)\calvbar_{e,m}(\zbar)
\ee
where
\be
\begin{split}
\calv_{e,m}(z) =e^{i\left(\frac{e}{R} + \frac{mR}{2}\right)\phi(z)}\\
\calvbar_{e,m}(\zbar) =e^{i\left(\frac{e}{R} - \frac{mR}{2}\right)\phibar(\zbar)}
\end{split}
\ee
of conformal dimension:
\be
(\Delta_{e,m},\Deltabar_{e,m}) = \Big(\shalf\big(\sfrac{e}{R}+\sfrac{mR}{2}\big)^2,\shalf\big(\sfrac{e}{R}-\sfrac{mR}{2}\big)^2\Big)
\ee
The OPE's between the chiral parts of the vertex operators are given by:
\be
\calv_{e,m}(z)\calv_{e',m'}(0)\sim z^{\left(\frac{e}{R}+\frac{mR}{2}\right)\left(\frac{e'}{R}+\frac{m'R}{2}\right)}\calv_{e+e',m+m'}(z) 
\ee

For the special value $R=1$, the compact boson is equivalent to a Dirac fermion. The fermion field has $(\Delta,\Deltabar)=(\half,\half)$ and is given, in bosonic language, by $\calo_{1,0}$. In Ref.\cite{Azeyanagi:2007bj} the fermionic twist fields are identified as:
\be
\sigma_k = \calo_{0,\frac{2k}{n}}
\ee
for $k=-\frac{n-1}{2},\cdots \frac{n-1}{2}$. These operators have $(\Delta,\Deltabar)=(\frac{k^2}{2n^2},\frac{k^2}{2n^2})$. Since the winding number $m$ for these operators is not an integer, they are not included in the set of local operators of the theory. This is expected, since twist operators are mutually non-local with the physical fields of the theory. One has the OPE's:
\be
\calo_{0,\frac{2k}{n}}(z,\zbar)\,\calv_{1,0}(w)\sim (z-w)^{\frac{k}{n}},\quad
\calo_{0,\frac{2k}{n}}(z,\zbar)\,\calvbar_{1,0}(\wbar)\sim (\zbar-\wbar)^{-\frac{k}{n}}
\ee
as desired. Now we only need to compute:
\be
\llangle \calo_{0,\frac{2k}{n}}(z,\zbar)\calo_{0,-\frac{2k}{n}}(0)\rrangle
\ee
For this we use the general result (see for example Ref.\cite{DiFrancesco:1997nk}):
\be
\begin{split}
\llangle \calo_{e,m}(z,\zbar)\calo_{-e,-m}(0)\rrangle&=\left|\frac{\theta_1'(0|\tau)}{\theta_1(\frac{\ell}{L}|\tau)}\right|^{4\Delta_{e,m}}\times\\
&\qquad \frac{1}{|\eta(\tau)|^2}\sum_{e',m'}q^{2\Delta_{e',m'}}
\qbar^{2\Deltabar_{e',m'}}\,e^{4\pi i(\alpha_{e',m'}\alpha_{e,m}z-\alphabar_{e',m'}\alphabar_{e,m}\zbar)}
\label{genres}
\end{split}
\ee
With $(e,m) = (0,\frac{2k}{n})$ and $R=1$, this reduces to:
\be
\begin{split}
\llangle \calo_{0,\frac{2k}{n}}(z,\zbar)\calo_{0,-\frac{2k}{n}}(0)\rrangle&=\left|\frac{\theta_1'(0|\tau)}{\theta_1(\frac{\ell}{L}|\tau)}\right|^{\frac{2k^2}{n^2}} \frac{1}{|\eta(\tau)|^2}
\sum_{e,m}q^{2\Delta_{e,m}}\qbar^{2\Deltabar_{e,m}}\,e^{4\pi i\frac{k\ell}{nL}e}\\
&=\left|\frac{\theta_1'(0|\tau)}{\theta_1(\frac{\ell}{L}|\tau)}\right|^{\frac{2k^2}{n^2}}\times \half\frac{\sum_{\nu=1}^4|\theta_\nu(\frac{k\ell}{nL}|\tau)|^2}{|\eta(\tau)|^2}\end{split}
\label{bossum}
\ee
An important point to emphasise here is that in this approach one has to use free bosons to represent {\em fermionic} twist fields. These do not introduce a phase on the boson, instead they {\em shift} the free boson $\phi$ appearing in $\psi=e^{i\phi}$ by sending $\phi\to \phi + \frac{2\pi k}{n}$ which is the $k/n$'th multiple of its natural period $2\pi$ (because $R=1$). Correspondingly one has $\phibar\to \phibar - \frac{2\pi k}{n}$ (here $\phi,\phibar$ are the holomorphic and anti-holomorphic parts of the scalar field respectively). The bosonic representation is used only as a tool to obtain the correlators of the fermion twist fields. 

At this point, we need to decide how to take the product over replicas. The most straightforward way would be to just take the product of the above result over all $k$. In this way the spin structures are summed over before we carry out replication. Hence the spin structure on one replica of the torus is uncorrelated with that on any other, leading to what we call the ``uncorrelated replica partition function'' $Z_n^{\rm u}$:
\be
\begin{split}
Z_n^{\rm u}(L,\beta;\ell)&=\prod_{k=-\frac{n-1}{2}}^{\frac{n-1}{2}}\llangle \calo_{0,\frac{2k}{n}}(z,\zbar)\calo_{0,-\frac{2k}{n}}(0)\rrangle\\
&= \left|\frac{\theta_1'(0|\tau)}{\theta_1(\frac{\ell}{L}|\tau)}\right|^{\frac16(n-\frac{1}{n})}\prod_{k=-\frac{n-1}{2}}^{\frac{n-1}{2}}\half\frac{\sum_{\nu=1}^4|\theta_\nu(\frac{k\ell}{nL}|\tau)|^2}{|\eta(\tau)|^2}
 \end{split}
\label{uncorr}
\ee
There is another way that one can take the product. First isolate the contribution to \eref{bossum} corresponding to a given spin structure $\nu\in 1,2,3,4$. This simply corresponds to picking out the function $\theta_\nu$. Next, take the product of this term over all replicas and finally sum over spin structures. In this way the spin structure on each replica of the torus is the same. We call this the ``correlated replica partition function'' $Z_n^{\rm c}$:
\be
Z_n^{\rm c}(L,\beta;\ell)= \half \left|\frac{\theta_1'(0|\tau)}{\theta_1(\frac{\ell}{L}|\tau)}\right|^{\frac16(n-\frac{1}{n})}\sum_{\nu=1}^4 \prod_{k=-\frac{n-1}{2}}^{\frac{n-1}{2}}\frac{|\theta_\nu(\frac{k\ell}{nL}|\tau)|^2}{|\eta(\tau)|^2}
\label{corr}
\ee
Notice that the two types of replica partition functions coincide at $n=1$:
\be
Z_1^{\rm u}= Z_1^{\rm c}=Z_1=\half\frac{\sum_{\nu=1}^4|\theta_\nu(0|\tau)|^2}{|\eta(\tau)|^2}
\ee
which is the ordinary modular-invariant partition function of a Dirac fermion (the contribution from $\nu=1$ actually vanishes in this case). We also observe that as $\ell\to 0$ (and generic  $n$), the two types of partition function are quite different:
\be
\begin{split}
Z_n^{\rm u}(L,\beta;\ell\to 0)&\sim \left(\frac{\ell}{L}\right)^{-\frac16\left(n-\frac{1}{n}\right)}
\left(\half\frac{\sum_{\nu=1}^4 |\theta_\nu(0|\tau)|^2}{|\eta(\tau)|^2}\right)^n\\
Z_n^{\rm c}(L,\beta;\ell\to 0)&\sim \left(\frac{\ell}{L}\right)^{-\frac16\left(n-\frac{1}{n}\right)}
\half\frac{\sum_{\nu=1}^4 |\theta_\nu(0|\tau)|^{2n}}{|\eta(\tau)|^{2n}}
\end{split}
\label{smallell}
\ee
The second factors in the two cases are the ordinary partition functions of $n$ Dirac fermions with, respectively, uncorrelated and correlated spin structures\cite{Elitzur:1986ye}.

In principle we can consider several more variations in which we correlate and sum over the spin structures on various subsets of the $n$ replicas, and let them remain uncorrelated across replicas. Such quantities are also modular covariant as one can verify. For the moment we restrict ourselves to these two ``extreme'' cases and will comment about the intermediate cases later on. We could also consider variations in which we sum over a subset of the spin structures in a correlated fashion and the remaining ones in an uncorrelated fashion. However, for modular covariance it is essential to sum over all spin structures in a symmetric way. Hence this possibility is not covariant and we will not consider it further.

Returning to the two replica partitions described above, we can define two \Renyi entropies:
\be
\begin{split}
S_n^{\rm u}&= \lim_{n\to 1}\frac{1}{1-n}\log\frac{Z_n^{\rm u}}{(Z_1)^n}\\
S_n^{\rm c}&= \lim_{n\to 1}\frac{1}{1-n}\log\frac{Z_n^{\rm c}}{(Z_1)^n}\\
\label{tworenyi}
\end{split}
\ee
The denominators are the same because, as we pointed out earlier, the two types of partition functions coincide at $n=1$.

Clearly we would like to know which one of these (if any) is the correct \Renyi entropy of the modular-invariant free Dirac fermion theory. Before addressing this question, which turns out to be quite subtle, let us establish a few properties of the two quantities $Z_n^{\rm u}$ and $Z_n^{\rm c}$. First of all, they can be compared to the non-modular-invariant result of Ref.\cite{Azeyanagi:2007bj} for the replica partition function:
\be
Z_n^{(\nu)}=\left|\frac{\theta_1'(0|\tau)}{\theta_1(\frac{\ell}{L}|\tau)}\right|^{\frac16(n-\frac{1}{n})}\prod_{k=-\frac{n-1}{2}}^{\frac{n-1}{2}}\frac{|\theta_\nu(\frac{k\ell}{nL}|\tau)|^2}{|\eta(\tau)|^2}
\label{nonmod}
\ee
and the corresponding \Renyi entropy:
\be
S_n^{(\nu)}=\frac{1}{1-n}\log \frac{Z_n^{(\nu)}}{(Z_1^{(\nu)})^n}
\label{renyiss}
\ee
This makes sense only for the spin structures $2,3,4$, since $Z_1^{(1)}$ vanishes and the \Renyi entropies in this spin structure are correspondingly ill-defined. Notice that the expressions in \eref{tworenyi} are not made from linear combinations of the ratios in \eref{renyiss}.

Let us now check the modular transformation properties of the expressions in \eref{uncorr} and \eref{corr}. This transformation exchanges the two cycles of the torus. We also have a branch cut along the horizontal axis, which under this transformation becomes a branch cut along the vertical axis. Thus the transformation acts as $\beta\leftrightarrow L$ and $\ell\to i\ell$, where we have used the identification $\tau=i\tau_2=i\frac{\beta}{L}$ and $z=\frac{\ell}{L}$. This permits us to use:
\be
\theta_{\alpha\beta}\left(\frac{z}{\tau}\Big|-\frac{1}{\tau}\right)= (-i)^{\alpha\beta}(-i\tau)^\half e^\frac{i\pi z^2}{\tau}\theta_{\beta\alpha}(z,\tau)
\ee
with the usual dictionary: $\theta_{11}\to-\theta_1, \theta_{10}\to \theta_2, \theta_{00}\to\theta_3,\theta_{01}\to\theta_4$. Applying this to the second term in \eref{uncorr} or \eref{corr}, one finds that even though the two expressions are different, they pick up the same multiplicative factor:
\be
e^{\frac{i\pi}{6\tau}\left(n-\frac{1}{n}\right)\left(\frac{\ell}{L}\right)^2}
\ee
From the  first term, which is common to both expressions, we get a corresponding factor:
\be
e^{-\frac{i\pi}{6\tau}\left(n-\frac{1}{n}\right)\left(\frac{\ell}{L}\right)^2}
\ee
which cancels the previous one, as well as a factor:
\be
|\tau|^{{\frac16}\left(n-\frac{1}{n}\right)}
\ee
Thus at the end, one has:
\be
Z_n^{\rm u,c}(\beta,L;i\ell)=\left(\frac{\beta}{L}\right)^{\frac16\left(n-\frac{1}{n}\right)} Z_n^{\rm u,c}(L,\beta;\ell)
\ee
We see that even after summing over spin structures, the replica partition functions are not modular invariant, but rather modular covariant -- they acquire a multiplicative pre-factor. This factor vanishes at $n=1$, so $Z_1$ is indeed modular invariant as it must be. 

The origin of the pre-factor lies in the fact that the replica partition function $Z_n$ is not a partition function at all, but rather a correlation function -- of twist fields. On the torus, the two-point function of a field $\phi(z,\zbar)$ of conformal dimension $(\Delta,\Deltabar)$ transforms as (for this and other standard results in CFT, see for example Ref.\cite{DiFrancesco:1997nk}):
\be
\left\langle \phi\left(\sfrac{z}{c\tau+d},\sfrac{\zbar}{c\bar\tau+d}\right)\,\phi(0,0)\right\rangle_{\frac{a\tau+b}{c\tau+d}}
=|c\tau+d|^{2(\Delta+\Deltabar)}\langle \phi(z,\zbar)\phi(0)\rangle_\tau
\ee
From this it is easy to verify that the product of twist-field correlators should transform precisely as we have found above. Indeed, since the dimensions $(\Delta_k,\Deltabar_k)$ of twist fields satisfy the universal relation:
\be
\sum_k\Delta_k=\frac{c}{24}\left(n-\frac{1}{n}\right)
\ee
one expects that for every CFT of central charge $c$ the analogous result will hold with:
\be
Z_n(\beta,L;i\ell)=\left(\frac{\beta}{L}\right)^{\frac{c}{6}\left(n-\frac{1}{n}\right)}Z_n(L,\beta;\ell)
\label{reptrans}
\ee
In what follows, we will explicitly verify this in all known cases.

One still has to physically understand the impact of this modular covariance on the \Renyi entropy $S_n$. Under \eref{reptrans}, one finds that $S_n$ that transforms as:
\be
S_n(\beta,L)=-\frac{c(n+1)}{6n}\log\frac{\beta}{L}+S_n(L,\beta)
\ee
The additive term survives in the limit $n\to 1$, therefore it even appears in the entanglement entropy. Now we know that the entanglement entropy always contains a non-universal additive constant (see for example Ref.\cite{Calabrese:2009qy}), and one can take the point of view that this constant simply changes under modular transformations. Such an interpretation is supported by the fact that the additive term is independent of the length $\ell$ of the entangling interval.

Alternatively, we can modify the prescription for computing the replica partition functions (in both uncorrelated and correlated cases) by multiplying them by an overall factor that renders them modular invariant. This is done as follows (importantly the factor is independent of $\ell$):
\be
{\tilde Z}_n = \left(\frac{\beta}{L}\right)^{\frac{c}{12}\left(n-\frac{1}{n}\right)} Z_n
\label{zedenconj}
\ee
We have put in a dependence on $c$, the central charge, which is equal to 1 in the present example. One expects that the replica partition function for every CFT, after multiplication by the above factor, will be modular invariant. From the discussion above, such a modification of $Z_n$ can be achieved by re-normalising all the twist fields as:
\be
\sigma_k\to \left|\frac{\beta}{L}\right|^{\Delta_k}\sigma_k
\ee
using the fact that $\sum_k \Delta_k=\frac{c}{24}(n-\frac{1}{n})$.

Returning now to the Dirac fermion, the two \Renyi entropies, defined as:
\be
\tS^{\rm u,c}_n = \frac{1}{1-n}\log \frac{\tZ^{\rm u,c}_n}{(Z_1)^n}
\ee
will satisfy:
\be
\tS^{\rm u,c}_n(\beta,L;i\ell) = \tS^{\rm u,c}_n(L,\beta;\ell)
\ee
It follows that the entanglement entropies obtained by taking $n\to 1$ in the above will also be modular invariant. 

In the following subsection we address the question of which one of $S^{\rm u}_n,S^{\rm c}_n$ is the correct \Renyi entropy of the theory. From now on, we will not be very careful to insert the prefactor above or to distinguish between modular covariant and modular invariant expressions at each stage, since this insertion is $\ell$-independent and can always be carried out at the end.

\section{Relation to thermodynamic entropy}

In Ref.\cite{Azeyanagi:2007bj}, the result \eref{nonmod} was used to provide evidence for the conjecture, based on holography, that the entanglement entropy satisfies:
\be
\lim_{\ell\to 0}\Big(S_E(L-\ell)-S_E(\ell)\Big)=S_{\rm th}
\label{entherm}
\ee
where $S$ is the thermodynamic entropy of the theory:
\be
S_{\rm th} = \beta^2\frac{\del}{\del\beta}\left(-\frac{1}{\beta}\log Z_1\right)
\ee
We will use this as a guide to determine which of our modular-invariant definitions of the replica partition function, $Z_n^{\rm u}$ or $Z_n^{\rm c}$, is correct. As emphasised in Refs.\cite{Cardy:2014jwa,Chen:2014ehg}, we can simply use the low-temperature expansion for this purpose. Note that the overall factor that we removed from the replica partition function to render it modular-invariant is independent of the interval size $\ell$ and therefore drops out of \eref{entherm}.

Consider now the small-$\ell$ behaviour of the candidate \Renyi entropies. From \eref{smallell} we see that in this limit, $Z_n^{\rm u}\to (Z_1)^n$ but $Z_n^{\rm c}\not\to (Z_1)^n$ (upto an overall power of $\ell$). It is a physical requirement that for small intervals (equivalent to a very large spatial size $L$) the replica partition function should indeed tend to $(Z_1)^n$ times the given power of $\ell$. On this criterion, the correct \Renyi entropy for small $\ell$ should be $S_n^{\rm u}$ and not $S_n^{\rm c}$. 

Next consider the large interval limit, $\ell \to L$. In this case, it has been predicted on general grounds \cite{Cardy:2014jwa,Chen:2014ehg} that the replica partition function should tend to $Z_1(n\tau)$, i.e. the ordinary partition function on a torus with $n$ times the original modular parameter, apart from the same power of $\ell$ as in the $\ell\to 0$ case. Moreover it has been shown that this behaviour (together with the correct $\ell\to 0$ behaviour) ensures that \eref{entherm} holds. Let us subject our $Z_n^{\rm u}$ and $Z_n^{\rm c}$ to this test. We start with the latter. From \eref{corr} one sees that:
\be
Z_n^{\rm c}(\ell\to L) = \half\left(\frac{\ell}{L}\right)^{-\frac16\left(n-\frac{1}{n}\right)}\sum_{\nu=1}^4 \prod_{k=-\frac{n-1}{2}}^{\frac{n-1}{2}}\frac{|\theta_\nu(\frac{k}{n}|\tau)|^2}{|\eta(\tau)|^2}
\ee
We may now apply an identity involving $\theta$-functions after fractional shifts\cite{WolframTheta:}:
\be
\prod_{k=-\frac{n-1}{2}}^{\frac{n-1}{2}}\Big|\theta_\nu\Big(\frac{k}{n}-z\Big|\tau\Big)\Big|=
\left(\prod_{p=1}^\infty\left|\frac{(1-q^{2p})^n}{1-q^{2pn}}\right|\right)
\big|\theta_\nu(nz|n\tau)\big|
\label{nangle}
\ee
This is easily derived using the product representation of $\theta$-functions.
It is important that the above relation is true for the three spin structures $\nu=2,3,4$ with no relative factors. The fourth spin structure $\nu=1$ does not matter because we have to put $z=0$ in the above equation, and $\theta_1(0|\tau)=0$. Using the above relation, we find:
\be
\begin{split}
Z_n^{\rm c}(\ell\to L) &= \half\left(\frac{\ell}{L}\right)^{-\frac16\left(n-\frac{1}{n}\right)}\sum_{\nu=1}^4
\frac{|\theta_\nu(0|n\tau)|^2}{|\eta(n\tau)|^2}\\
&=\left(\frac{\ell}{L}\right)^{-\frac16\left(n-\frac{1}{n}\right)}Z_1(n\tau)
\end{split}
\ee
Thus we see that for a large interval, $Z_n^{\rm c}$ becomes equal to $Z_1(n\tau)$ upto the standard power of $\ell$, as desired. However, if we repeat the analysis for $Z_n^{\rm u}$ then this is not the case.  With this function we would find cross-terms between $\theta$-functions of different spin structures, for which an identity like \eref{nangle} does not hold. We thus have clear evidence that as $\ell\to L$, the correct replica partition function should be $Z_n^{\rm c}$ and not $Z_n^{\rm u}$.  

Hence we are led to propose that the correct modular-invariant partition function for the free Dirac fermion theory is $Z_n^{\rm u}$ at small $\ell$ and $Z_n^{\rm c}$ at large $\ell$. 
Let us now verify that this proposal precisely satisfies \eref{entherm}. We have:
\be
\begin{split}
\lim_{\ell\to 0}\Big(S_E(L-\ell)-S_E(\ell)\Big) &=
\lim_{n\to 1}\frac{1}{1-n}\log\left(\frac{Z_1(n\tau)}{(Z_1(\tau))^n}\right)\\
& = \log Z_1\left(\frac{\beta}{L}\right) - \frac{\beta}{L}\frac{Z'(\frac{\beta}{L})}{Z(\frac{\beta}{L})}\\
&=S_{\rm th}
\end{split}
\ee

Thus our proposal satisfies the relation to thermodynamic entropy. Notably, the final steps work out just as in the general proof of Ref.\cite{Chen:2014ehg} (see their Eq.(11)) but here we have used explicit $\theta$-function identities to obtain them. The fact that spin structures are uncorrelated/correlated for small/large intervals is a nice mathematical verification, in a special class of models, of the general principles put forward in Refs.\cite{Cardy:2014jwa,Chen:2014ehg}. The key conclusion for us is that neither $S_n^{\rm u}$ nor $S_n^{\rm c}$ can be the correct \Renyi entropy for all values of $\ell$. This naturally raises the question of what is the general answer that interpolates between these two limits, a point to which we will return at the end.

As noted above, in addition to the fully correlated and fully uncorrelated replica partition functions, one can write down other candidate quantities in which the spin structures are correlated within subsets of the total set of replicas and uncorrelated among different subsets. It is straightforward to check that such quantities do not behave correctly in both the limits considered here, in other words as $\ell\to0$ they do not give $(Z_1)^n$ and as $\ell\to L$ they do not give $Z_1(n\beta)$. They are therefore not useful candidates to describe the limiting behaviour of the \Renyi/entanglement entropies. However, as we will suggest in the concluding section, they could play a role for intermediate values of $\ell$.

\section{Compact bosons}

The replica trick can be applied for free bosons as well. Recall that the partition function for a single free boson at radius $R$ is:
\be
\begin{split}
Z_1(R)&=\frac{1}{|\eta(\tau)|^2}\sum_{e,m}q^{\left(\frac{e}{R}+\frac{mR}{2}\right)^2}\qbar^{\left(\frac{e}{R}-\frac{mR}{2}\right)^2}\\
&=\frac{1}{|\eta(\tau)|^2}\sum_e q^{\frac{2e^2}{R^2}}\sum_m q^{\frac{m^2R^2}{2}}\\
\end{split}
\label{zoner}
\ee
where in the second line we have specialised to real $q=e^{-\pi\tau_2}$.

For the replica partition function, one considers twist fields $\cT_k$ with $k=0,1,\cdots,n-1$ satisfying:
\be
\cT_k(z,\zbar)\phi(w)\sim (z-w)^{\frac{k}{n}},\quad \cT_k(z,\zbar)\phibar(\wbar)\sim (\zbar-\wbar)^{-\frac{k}{n}}
\label{bostwist}
\ee
and one has:
\be
Z_n =  \prod_{k=0}^{n-1}
\llangle \cT_k(z,\zbar)\cT_{-k}(0,0)\rrangle\\
\ee
An important part of the computation of this quantity was carried out in Ref.\cite{Datta:2013hba} for a pair of free bosons compactified on a square torus of size $R$. The result is the product of a quantum and a classical part. It is rather implicit, involving integrals of products of fractional powers of $\theta$-functions which appear in the construction of cut differentials on the torus. Unfortunately the classical part of their answer, which carries all the $R$-dependence, is not invariant under $R\to \frac{\alpha'}{R}$ so it does not satisfy T-duality. In Ref.\cite{Chen:2015cna} this classical part has been corrected and takes a different form that possesses manifest T-duality. 

Accordingly we will analyse the replica partition function obtained in the latter reference and compare it with our proposal regarding modular invariance. We adapt it to our notation, which includes setting $\alpha'=2$, and to the case of a single free boson. The latter just involves taking the square  root of their answer, which is particularly simple since it is presented as a perfect square. We write the answer as a product of four factors:
\be
Z_n(R) = Z_n^{(1)} Z_n^{(2)}\,Z_n^{(3)}\!(R)\,Z_n^{(3)}\!\!\left(\frac{2}{R}\right) 
\label{znr}
\ee
Only the last two factors, which form the classical part of the answer, depend on the compactification radius $R$. T-duality invariance is already manifest at this stage.
Here,
\be
\begin{split}
Z^{(1)} &= \frac{1}{|\eta(\tau)|^{2n}}\prod_{k=0}^{n-1}\frac{1}{|W_1^1(k,n;\sfrac{\ell}{L}|\tau)|}\\
Z^{(2)} &= \left|\frac{\theta_1'(0|\tau)}{\theta_1(\frac{\ell}{L}|\tau)}\right|^{\frac16\left(n-\frac{1}{n}\right)}\\
Z^{(3)}\!(R) &= \sum_{m_j}\exp\left(-\frac{\pi R^2}{2n}\sum_{k=0}^{n-1}\left|\frac{W_2^2(k,n)}{W_1^1(k,n)}\right|
\sum_{j,j'=0}^{n-1}\left[\cos 2\pi(j-j')\frac{k}{n}\right]m_j m_{j'}\right)\\
\end{split}
\ee
Here $W_1^1$ and $W_2^2$ are integrals of the cut differentials over the different periods of the torus:
\be
\begin{split}
W_1^1(k,n;\sfrac{\ell}{L}|\tau) & = \int_0^1 dz~ \theta_1(z|\tau)^{-\left(1-\frac{k}{n}\right)}
\theta_1\left(z-\sfrac{\ell}{L}|\tau\right)^{-\frac{k}{n}}\theta_1\left(z-\sfrac{k\ell}{nL}|\tau\right)\\
W_2^2(k,n;\sfrac{\ell}{L}|\tau) & = \int_0^{\bar\tau} d\zbar~ \bar\theta_1(\zbar|\tau)^{-\frac{k}{n}}
\bar\theta_1\left(\zbar-\sfrac{\ell}{L}|\tau\right)^{-\left(1-\frac{k}{n}\right)}
\bar\theta_1\left(\zbar-\left(1-\sfrac{k}{n}\right)\sfrac{\ell}{L}|\tau\right)
\end{split}
\ee
The answer \eref{znr} is normalised such that $Z_1$ is indeed the usual partition function of the theory. To check this, observe that for $n=1,k=0$ one has $W_1^1=1, W_2^2=-i\tau_2$. Hence:
\be
\begin{split}
Z^{(1)} &= \frac{1}{|\eta(\tau)|^{2}}\\
Z^{(2)} &= 1\\
Z^{(3)}\!(R) &= \sum_{m}\exp\left(-\frac{\pi R^2}{2}\tau_2 m^2\right)=\sum_m q^{\frac{m^2R^2}{2}}\\
Z^{(3)}\!\left(\frac{2}{R}\right) &= \sum_{e}\exp\left(-\frac{2\pi}{R^2}\tau_2 e^2\right)=\sum_e q^{\frac{2e^2}{R^2}}\\
\end{split}
\ee
Agreement with \eref{zoner} is immediate.

We would now like to investigate the modular transformation of \eref{znr}. To this end, we note the following results:
\be
\begin{split}
\eta\Big(-\frac{1}{\tau}\Big) &=(-i\tau)^\half \eta(\tau)\\ 
W_1^1\Big(k,n;\sfrac{i\ell}{\beta}|-\sfrac{1}{\tau}\Big) &= \frac{1}{\tau}e^{-\frac{i\pi \ell^2}{L^2\tau}\frac{k}{n}\left(1-\frac{k}{n}\right)}W_2^2(k,n;\sfrac{\ell}{L}|\tau)\\
\frac{\theta_1'\!\left(0|-\frac{1}{\tau}\right)}{\theta_1\!\left(\frac{z}{\tau}|-\frac{1}{\tau}\right)} &=
i\tau e^{-\frac{i\pi z^2}{\tau}}
\frac{\theta_1'(0|\tau)}{\theta_1(z|{\tau})} \end{split}
\ee
The first and third of these are well-known, while the second was shown in Eq.(B.41) of Ref.\cite{Datta:2013hba} (unlike the very subtle issue of the $q$-expansion of $W_2^2$, this result is straightforward to derive).  From the above, it follows that under a modular transformation,
\be
\left|\frac{W_2^2(k,n)}{W_1^1(k,n)}\right| \to \left|\frac{W_1^1(k,n)}{W_2^2(k,n)}\right|
\ee
Following this with a multi-variable Poisson resummation as in Ref.\cite{Chen:2015cna}, we find that:
\be
\begin{split}
Z^{(3)}\Big(R;\frac{z}{\tau}\Big|-\frac{1}{\tau}\Big) &= \frac{2^{\frac{n}{2}}}{R^{n}}\left(\prod_{k=0}^{n-1}\left|\frac{W_2^2(k,n)}{W_1^1(k,n)}\right|^\half\right)
Z^{(3)}\Big(\frac{2}{R};z\Big|\tau\Big)\\
Z^{(3)}\Big(\frac{2}{R};\frac{z}{\tau}\Big|-\frac{1}{\tau}\Big) &= \frac{R^{n}}{2^\frac{n}{2}}\left(\prod_{k=0}^{n-1}\left|\frac{W_2^2(k,n)}{W_1^1(k,n)}\right|^\half\right)
Z^{(3)}\Big(R;z\Big|\tau\Big)
\end{split}
\ee
Thus the product transforms as:
\be
Z^{(3)}\!(R)\,Z^{(3)}\!\!\left(\frac{2}{R}\right)
\to \left(\prod_{k=0}^{n-1}\left|\frac{W_2^2(k,n)}{W_1^1(k,n)}\right|\right)
Z^{(3)}\!(R)\,Z^{(3)}\!\!\left(\frac{2}{R}\right)
\ee
Putting everything together, we find that:
\be
Z_n\!\left(R;\frac{z}{\tau}\Big|-\frac{1}{\tau}\right)=|\tau|^{\frac{1}{6}\left(n-\frac{1}{n}\right)}
Z_n(R;z|\tau)
\ee
Thus, it is modular covariant with the expected prefactor (see \eref{reptrans}). 

The general expression for \Renyi entropy \eref{znr} is rather complicated and it has proved difficult to extract explicit results from it. However, in Ref.\cite{Chen:2015cna} the limits as $\ell\to 0$ and $\ell\to L$ have been extracted and the results correspond (upto the usual powers of $\ell$) to $(Z_1(\tau))^n$ and $Z_1(n\tau)$ respectively, where by $Z_1$ we mean the expression in \eref{zoner}. This of course depends on $R$, and for $R=1$ it coincides with the free Dirac fermion partition function. It follows from this and our previous analysis that our fermion replica partition function, in the regimes where we know it (small and large $\ell$) coincides precisely with that of the boson theory at $R=1$. It is amusing that the boson theory ``knows'' about the switchover from uncorrelated to correlated replicas within the single, though complicated, function in \eref{znr}. We expect there to be a similar fermion partition function that interpolates between the small and large-$\ell$ regimes. It should somehow involve an admixture of correlated/uncorrelated replicas, as we discuss in the last section.

\section{Other free-fermion CFT's}

It is interesting to try and extend the previous calculation to free-field CFT's that are not direct sums of the Dirac CFT. Recall that in the above calculation we had to use both the fermionic formulation of the Dirac CFT (to identify the right free field to replicate) and also the bosonic formulation (to identify the twist field and compute its correlation function). For multiple Dirac fermions with correlated spin structures, the partition function is not simply a power of the single-Dirac partition function\footnote{In this part of the discussion there is some potential for confusion: the multiple fermions can have correlated/uncorrelated spin structures, which is part of the definition of the theory, while the replica partition function can also have correlated/uncorrelated spin structures across replicas as was discussed in previous sections.}. Correspondingly, the bosonic dual is not made up of copies of the $R=1$ free boson. Indeed in bosonic language it is not the product of any number of independent bosons but rather a set of bosons compactified on the weight lattice of the algebra Spin(2d) with a specific background $B$-field turned on\cite{Elitzur:1986ye}. We start with the relatively simple case of Spin(4)$\sim$ SU(2)$\times$ SU(2) and then go on to a general Spin(2d) lattice.
Calculation of the \Renyi entropy will again require use of both the fermion and boson descriptions of the theory. In each case, the twist field must be constructed using lattice vertex operators and their correlation functions computed in the theory of lattice bosons. We will again find answers for the replica partition function at small and large $\ell$ and verify their modular properties.

\subsection{Two correlated Dirac fermions}

The theory of two Dirac fermions with correlated spin structures has the ordinary partition function\cite{Elitzur:1986ye}:
\be
Z_1[2]= \half \sum_{\nu=2,3,4}\frac{|\theta_\nu(0|\tau)|^4}{|\eta(\tau)|^4}
\ee
We would like to find its replica partition function. Let us examine the fundamental fields and physical operators in this theory. We have two Dirac fermions that, following the notation of Ref.\cite{DiFrancesco:1997nk}, we call $D_1(z)$ and $D_2(z)$. By definition they possess hermitian conjugates $D_i^\dagger(z)$, $i=1,2$, that are also holomorphic. Combining with their anti-holomorphic counterparts, we have 16 physical operators of dimension $(\half,\half)$, namely:
\be
D_i\Dbar_j, D^\dagger_i \Dbar_j, D_i \Dbar^\dagger_j,D_i^\dagger \Dbar^\dagger_j
\ee
where $i.j$ run independently over $1,2$ (had the spin structures of $D_1,D_2$ been uncorrelated we would only have had the subset with $i=j$). This $c=2$ theory can be bosonised into two compact bosons $\phi_1,\phi_2$ that are orthogonal to each other and compactified at the self-dual radius, which in our conventions is $R=\sqrt2$. Thus their periodicity is:
\be
\phi_i \to \phi_i + 2\sqrt2 \pi
\ee
If we denote the two left-moving bosons as $\phi_i(z)$ and the right-moving ones as $\phibar_i(\zbar)$, then the allowed vertex operators in this theory are:
\be
e^{\frac{i}{\sqrt2}\left(e_j+m_j\right)\phi_j(z)}e^{\frac{i}{\sqrt2}\left(e_k-m_k\right)\phibar_k(\zbar)}
\ee
where the indices $j,k$ are independently summed over $1,2$. 
In terms of these bosons, the free Dirac fields are represented as:
\be
D_1(z) = e^{\frac{i}{\sqrt2}\phi_1}e^{\frac{i}{\sqrt2}\phi_2},\quad
D_2(z) = e^{\frac{i}{\sqrt2}\phi_1}e^{-\frac{i}{\sqrt2}\phi_2}
\label{twodirac}
\ee
The Hermitian conjugates have the same expressions but with $i\to -i$, while the anti-holomorphic conjugates have $\phi_i\to \phibar_i$ without any change of sign on $i$. The 16 local operators obtained after combining left- and right-movers correspond to all pairs of orthogonal vectors of $({\rm length})^2=2$ in the unit 2d square lattice:
\be
\begin{split}
(e_i;m_i) &= \pm((1,1);(0,0)),~~\pm((1,-1);(0,0)),~~\pm((0,0);(1,1)),~~\pm((0,0);(1,-1)),\\
&\pm((1,0);(0,1)),~~\pm((1,0);(0,-1)),~~\pm((0,1);(1,0)),~~\pm((0,1);(-1,0))
\end{split}
\ee

In order to compute the replica partition function we need to find the twist fields which satisfy \eref{twact}. On inspecting \eref{twodirac}, we see that the monodromy induced by the twist field is:
\be
\begin{split}
&e^{\frac{i}{\sqrt2}\phi_1}e^{\frac{i}{\sqrt2}\phi_2}\to e^{\frac{2\pi ik}{n}}
e^{\frac{i}{\sqrt2}\phi_1}e^{\frac{i}{\sqrt2}\phi_2}\\
&e^{\frac{i}{\sqrt2}\phi_1}e^{-\frac{i}{\sqrt2}\phi_2}\to e^{\frac{2\pi ik}{n}}
e^{\frac{i}{\sqrt2}\phi_1}e^{-\frac{i}{\sqrt2}\phi_2}
\end{split}
\ee
Thus it must leave the boson $\phi_2$ inert, while it shifts $\phi_1$ by a fraction $\frac{k}{n}$ of its period:
\be
\phi_1\to \phi_1+\frac{2\sqrt2\pi k}{n}
\ee
We see that the twist field is:
\be
\sigma_k = e^{\sqrt2 i\frac{k}{n}\phi_1}e^{-\sqrt2 i\frac{k}{n}\phibar_1}
\ee
In other words, it corresponds to ${\cal O}^{(1)}_{0,\frac{2k}{n}}$. Remarkably this is the same operator that plays the role of the twist field in the $c=1$ theory of a single Dirac fermion -- however, in the present case it is evaluated at $R=\sqrt2$ rather than $R=1$. Due to the change of radius, it has conformal dimension $\Delta_{0,\frac{2k}{n}}=\frac{k^2}{n^2}$, which is twice that in the single-Dirac case. This ensures that $\sum_k \Delta_k=\frac{c}{24}\left(n-\frac{1}{n}\right)$ with $c=2$, as required. 

To calculate its correlator we may use \eref{genres} and make the appropriate change in radius, to get:
\be
\begin{split}
\llangle \calo_{0,\frac{2k}{n}}(z,\zbar)\calo_{0,-\frac{2k}{n}}(0)\rrangle&=\left|\frac{\theta_1'(0|\tau)}{\theta_1(\frac{\ell}{L}|\tau)}\right|^{\frac{4k^2}{n^2}} \frac{1}{|\eta(\tau)|^4}\times\\
&\sum_{e_1,m_1;e_2,m_2}q^{2(\Delta_{e_1,m_1}+\Delta_{e_2,m_2})}\qbar^{2(\Deltabar_{e_1,m_1}+\Deltabar_{e_2,m_2})}\,e^{4\pi i\frac{k\ell}{nL}e_1}\\
&=\left|\frac{\theta_1'(0|\tau)}{\theta_1(\frac{\ell}{L}|\tau)}\right|^{\frac{4k^2}{n^2}}\times \half\frac{\sum_{\nu=1}^4|\theta_\nu(\frac{k\ell}{nL}|\tau)|^4}{|\eta(\tau)|^4}
\end{split}
\label{bossumtwo}
\ee
The last line follows by the change of variables:
\be
\begin{split}
e_1&=\half(n_1+n_2+n_3+n_4),\quad m_1=\half(n_1+n_2-n_3-n_4)\\
e_2&=\half(n_1-n_2+n_3-n_4),\quad m_2=\half(n_1-n_2-n_3+n_4)\\
\end{split}
\ee
The $n_i$ can be all integers or all half-integers. Moreover, the sum of all the $n_i$ is constrained to be even. Upon implementing these two facts, the sum in \eref{bossumtwo} splits into four parts, one corresponding to each spin structure. In this way we recover the last line of \eref{bossumtwo}.

We now have to take the product over replicas. As before, we can do this after or before summing over spin structures. The results are then:
\be
\begin{split}
Z^{\rm u}_n &= \left|\frac{\theta_1'(0|\tau)}{\theta_1(\frac{\ell}{L}|\tau)}\right|^{\frac{1}{3}(n-\frac{1}{n})}
\prod_{k=-\frac{n-1}{2}}^{\frac{n-1}{2}}
\half \frac{\sum_{\nu=1}^4 
|\theta_\nu(\frac{k\ell}{nL}|\tau)|^4}{|\eta(\tau)|^{4}}\\
Z^{\rm c}_n &= \left|\frac{\theta_1'(0|\tau)}{\theta_1(\frac{\ell}{L}|\tau)}\right|^{\frac{1}{3}
(n-\frac{1}{n})}\half
\sum_{\nu=1}^4 \prod_{k=-\frac{n-1}{2}}^{\frac{n-1}{2}}
\frac{
|\theta_\nu(\frac{k\ell}{nL}|\tau)|^4}{|\eta(\tau)|^{4}}
\end{split}
\label{twodir}
\ee
We again propose that the first result gives the correct \Renyi entropy as $\ell\to 0$ while the second one is correct when $\ell\to L$. 

As emphasised at the beginning of this section, the above partition functions do not directly follow from the single Dirac case. The reason is that one needs a bosonic dual theory in order to write the twist field. For two Dirac fermions with correlated spin structures, the bosonic dual is  two copies of the $R=\sqrt2$ compact boson. Interestingly the twist field that we found was not symmetric in the two bosons, depending only on one of them. We will see below that this pattern repeats for multiple correlated fermions, dual to bosons on Spin(2d) weight lattices.  

It is straightforward to demonstrate that under modular transformations the above replica partition functions satisfy:
\be
Z^{\rm u,c}_n(\beta,L)=\frac{\beta}{L}^{\frac13\left(n-\frac{1}{n}\right)} Z^{\rm u,c}_n(L,\beta)
\ee
which precisely agrees with our expectation.

\subsection{Multiple correlated fermions and lattice bosons}

For $d\ge 3$, the theory of $d$ free Dirac fermions with correlated spin structures is dual to a specific compactification of $d$ free bosons on a target-space torus:
\be
T^c=R^d/\Gamma_d
\ee
where $\Gamma_d$ is the root lattice of Spin$(2d)$ (this can be achieved by starting with a rectangular torus and choosing a suitable constant metric and $B$-field). In this case the $d$ different bosons are not orthogonal to each other, while the fermions have correlated spin structures, so on both sides of the Bose-Fermi duality we are dealing with CFT's that are not direct sums of simpler ones. 

The ordinary partition function is well-known and has been derived directly within the free fermion representation as well as from the free boson picture\cite{Elitzur:1986ye}. It corresponds to $Z_1[d]$ in terms of the definition in \eref{zedonem}. We would like to construct the replica partition functions for these theories. As in the previous examples, this will be possible only by using features of both the fermionic and bosonic descriptions together.

The first step in doing this is to identify the $(\half,\half)$ operators  in terms of the bosons. As in the case of two Dirac fermions, we have the operators:
\be
D_p\Dbar_q, D^\dagger_p \Dbar_q, D_p \Dbar^\dagger_q,D_p^\dagger \Dbar^\dagger_q
\ee
where $p,q=1,2,\cdots d$. In the free boson theory, let $\Lambda_R$ be the root lattice and $\Lambda_W$ be the dual weight lattice. Then the vertex operators are:
\be
\cO_{w^i,\wbar^i}=e^{iw^i\phi_i}e^{i\wbar^i\phibar_i}
\ee
where $w^i,\wbar^i\in \Lambda_W$ and $w^i-\wbar^i\in \Lambda_R$. Elements of the weight lattice can be parametrised as:
\be
w^i = \frac{1}{\sqrt2}g^{ij}v_j,\qquad \wbar^i =\frac{1}{\sqrt2}g^{ij}\vbar_j
\label{emparam}
\ee
where $v_i,\vbar_i$ are integers and $g^{ij}$ is the inverse of $g_{ij}$ which is the half the Cartan matrix of Spin(2d).
For the difference to lie in the root lattice, we must require that $\frac{1}{\sqrt2}(v_i-\vbar_i)=\sqrt2 n_i$ where $n_i$ are integers.

The 2-point function of scalar fields is:
\be
\phi_i(z,\zbar)\phi(z',\zbar')=-g_{ij}\log|z-z'|^2
\ee
Hence the conformal dimension of the above operators is $(\Delta_{w^i},\Deltabar_{\wbar^i})=\half(g_{ij}w^iw^j,g_{ij}\wbar^i\wbar^j)$. Thus to reconstruct the fermion operators, we must look for pairs of points of unit length in the weight lattice that differ by an element of the root lattice. If $\valpha_i$ are the $d$ simple roots of Spin(2d) and $\vlambda^i$ are the fundamental weights then:
\be
\valpha_i\cdot \vlambda^j=\delta_i^{~j},\quad g_{ij}=\half\valpha_i\cdot \valpha_j,\quad g^{ij}=2\vlambda^i\cdot \vlambda^j
\ee
One also has the dual relation:
\be
(\vlambda^i)_p(\valpha_i)_q=\delta_{pq}
\label{dualrel}
\ee
which will be important in what follows. Here $p,q$ label the individual components of each vector, and the sum is over the vectors (not components).

For the conformal dimensions, we have:
\be
\Delta_{w^i}=\half g_{ij}w^iw^j = \frac14 g^{ij}v_iv_j
\ee
Therefore to find operators of dimension $(\half,\half)$ we must look for sets of integers $v_i$ for which:
\be
g^{ij}v_iv_j=2
\ee
From \eref{dualrel} it follows that:
\be
g^{ij}(\valpha_i)_p(\valpha_j)_q=2\delta_{pq}
\label{lamalph}
\ee 
hence the possible $v_i$ are given by:
\be
v^{(p)}_i=(\valpha_i)_p,\quad p=1,2,\cdots,d
\ee
For the anti-holomorphic part we start by picking it independently from the same set: $\vbar^{(q)}_i=(\valpha_i)_q$. However, we need to ensure that
$\frac{1}{\sqrt2}g^{ij}\left(v^{(p)}_j-\vbar^{(q)}_j\right) $ is $\sqrt2$ times an integer. This is guaranteed by the fact that
\be
\frac{1}{\sqrt2}g^{ij}\left(v^{(p)}_j-\vbar^{(q)}_j\right) 
= \frac{1}{\sqrt2}g^{ij}\big((\valpha_j)_p-(\valpha_j)_q\big)
=\sqrt2\big((\vlambda^j)_p-(\vlambda^j)_q\big)
\ee
which is indeed in the root lattice. We conclude that the fermion operators are given in bosonic language by:
\be
D_p(z)\Dbar_q(\zbar)\to \cO_{p,q}=e^{i w^{(p)}{}^i\phi_i(z)}e^{i\wbar^{(q)}{}^i\phibar_i(\zbar)}
\ee
where $w^{(p)}{}^i = \sqrt2 (\vlambda^i)_p$.

We can now look for the twist field, an operator $\sigma_k$ satisfying \eref{twact}. This induces a monodromy:
\be
\sigma_k: D_p(z)\to e^{\frac{2\pi ik}{n}}D_p(z)
\ee
corresponding to a shift:
\be
w^{(p)}{}^i\phi_i(z)\to w^{(p)}{}^i\phi_i(z)+\frac{2\pi k}{n}
\ee
This in turn will be induced by a shift $\phi_i\to  \phi_i + 2\pi \zeta^{(k)}_i$ where $\zeta^{(k)}_i$ is a constant vector satisfying:
\be
w^{(p)}{}^i\zeta^{(k)}_i = \frac{k}{n}
\ee
for all $p$. Recalling that the last weight of Spin(2d) is $\lambda^{(d)}=(\half,\half,\cdots,\half)$, 
we easily find that the shift is given by:
\be
\zeta^{(k)}_i = \frac{\sqrt 2 k}{n}(0,0,\cdots,0,1)
\ee
Thus the twist field only acts on the last scalar $\phi_d$. Indeed, it takes the form:
\be
\sigma_k = \cO_{\zeta^{(k)}{}^i,-\zeta^{(k)}{}^i}=e^{i\zeta^{(k)}{}^i\phi_i(z)}e^{-i\zeta^{(k)}{}^i\phibar_i(\zbar)}
\ee
where one must keep in mind that $\zeta^{(k)}{}^i=g^{ij}\zeta^{(k)}_j$. To check that this is correct, compute the OPE with $D_p(z)$ to find:
\be
\sigma_k(z)e^{iw^{(p)}{}^i\phi_i(z')}\sim (z-z')^{\zeta^{(k)}{}^i g_{ij}w^{(p)}{}^j}=(z-z')^{w^{(p)}{}^i\zeta^{(k)}{}_i}
=(z-z')^{\frac{k}{n}}
\ee
The crucial test of our twist field is whether it has the desired conformal dimension. We have:
\be
\Delta_k = \half g_{ij}\,\zeta^{(k)}{}^i\zeta^{(k)}{}^j=\half g^{ij}\zeta^{(k)}_i\zeta^{(k)}_j = \frac{k^2}{n^2}\,g^{dd}
\ee
Using $g^{dd}= 2\lambda^{(d)}\cdot \lambda^{(d)}=\frac{d}{2}$ we get $\Delta_k = \frac{dk^2}{2n^2}$ from which it follows that:
\be
\sum_{k=-\frac{n-1}{2}}^{k=\frac{n-1}{2}} \Delta_k = \frac{d}{24}\left(n-\frac{1}{n}\right)
\ee
as desired.

It remains to calculate the two-point function of each $\sigma_k$ and thereby the replica partition function.
Recall that the ordinary partition function for these theories is:
\be
\begin{split}
Z_1 &= \frac{1}{|\eta(\tau)|^{2d}}\sum_{\genfrac{}{}{0pt}{}{w,\wbar\in \Lambda_W}{w-\wbar\in\Lambda_R} }
q^{w^2}\qbar^{\,\wbar^2}\\
& = \half \frac{1}{|\eta(\tau)|^{2d}}\sum_{\nu=2,3,4}|\theta_\nu(0|\tau)|^{2d}
\end{split}
\ee
where, as usual, $w^2=g_{ij}w^i w^j$ and similarly for $\wbar^2$. The equivalence between the first and second lines of the above equation is the statement of Bose-Fermi equivalence between lattice bosons and multiple fermions with correlated spin structures. To demonstrate this, one starts with the observation that for Spin(2d), $\Lambda_W = \Lambda_R \oplus \Lambda_V \oplus \Lambda_S \oplus \Lambda_C$ where $\Lambda_{V,S,C}$ are the lattices obtained by shifting $\Lambda_R$ by a fundamental weight in the vector, spinor and conjugate spinor representations respectively. Now we split a generic vector $w\in \Lambda_W$ into two classes: those that lie is $\Lambda_R\cup\Lambda_V$ and those in $\Lambda_S\cup \Lambda_C$. In the former set we can write $w^i=\sqrt2\sum_p n_p(\vlambda^i)_p$ for arbitrary integers $n_p$, while in the latter set $w^i=\sqrt2\sum_p (n_p+\half)(\vlambda^i)_p$ where $n_p$ are again arbitrary integers. We similarly write $\wbar^i=\sqrt2\sum_p m_p(\vlambda^i)_p$ and $\wbar^i=\sqrt2\sum_p (m_p+\half)(\vlambda^i)_p$ in the two respective cases.  Finally, the restriction $w-\wbar\in \Lambda_R$ is implemented by inserting a projection operator that causes both $w$ and $\wbar$ to lie in $\Lambda_R$ or in $\Lambda_V$ (in the first set) or both to lie in $\Lambda_S$ or in $\Lambda_C$ (in the second set). In this way we end up with the four Jacobi $\theta$-functions $\theta_\nu(0|\tau),\nu=1,2,3,4$, of which $\theta_1$ vanishes for familiar reasons. 

For the un-normalised two-point function of twist fields, we find instead:
\be
\llangle \sigma_k(z,\zbar) \sigma_{-k}(0)\rrangle=\left|\frac{\theta_1'(0|\tau)}{\theta_1(\frac{\ell}{L}|\tau)}\right|^{\frac{2dk^2}{n^2}} \frac{1}{|\eta(\tau)|^{2d}}
\sum_{\genfrac{}{}{0pt}{}{w,\wbar\in \Lambda_W}{w-\wbar\in\Lambda_R} }q^{w^2}\qbar^{\,\wbar^2}
e^{2\pi i \frac{\ell}{L} g_{ij}(w^i+\wbar^i)\zeta^{(k)}{}^j}
\label{latsum}
\ee
Now we have 
\be
\begin{split}
g_{ij}(w^i+\wbar^i)\zeta^{(k)}{}^j &= (w^i+\wbar^i)\zeta^{(k)}_i=\frac{\sqrt2 k}{n}(w^d+\wbar^d)\\
& = \frac{k}{n}\sum_{p=1}^d(n_p+m_p),\quad w,\wbar\in \Lambda_R\cup\Lambda_V\\
& = \frac{k}{n}\sum_{p=1}^d(n_p+m_p-1),\quad w,\wbar\in \Lambda_S\cup\Lambda_C
\end{split}
\ee
It follows that:
\be
\begin{split}
\llangle \sigma_k(z,\zbar) \sigma_{-k}(0)\rrangle&=\half\left|\frac{\theta_1'(0|\tau)}{\theta_1(\frac{\ell}{L}|\tau)}\right|^{\frac{2dk^2}{n^2}} 
 \frac{\sum_{\nu=1}^4|\theta(\frac{k\ell}{nL}|\tau)|^{2d}}{|\eta(\tau)|^{2d}}
\end{split}
\ee
Taking the product over $k$ outside or inside the sum over spin structures gives us, respectively, the uncorrelated/correlated replica partition functions. As in the previous cases, under a modular transformation the replica partition function is covariant with the expected prefactor.

\section{Conclusions}

The principle of modular invariance has been consistently useful in understanding conformal field theory. Here we have applied it to the study of \Renyi entropies of free Dirac fermions and found that one can define more than one modular-covariant replica partition function. Our proposal is that the uncorrelated version, taking the product over replicas performed after the sum over spin structures, defines the correct \Renyi/entanglement entropy as $\ell\to 0$ while the correlated version, taking the product over replicas before summing over spin structures, defines the correct entropies as $\ell\to L$. 
We verified that in this way the holographically predicted relation to thermal entropy of Ref.\cite{Azeyanagi:2007bj} is satisfied. Comparing with results in the literature for free compact bosons, we showed that Bose-Fermi duality holds in the small and large-interval regimes.

It is clearly of importance to obtain the correct modular-covariant replica partition function at arbitrary values of $\ell$. For a generic positive integer $n$ one can write a variety of partition functions as follows. Consider the set $S=\{k_1,k_2,\cdots, k_n\}$ where $k_i=-\frac{n-1}{2}+i-1$. Suppose that for a fixed value $k_i$ and a fixed spin structure $\nu$, we have a replica partition function $Z_n^{(\nu)}(k_i)$. Partition $S$ into subsets $S_1,S_2,\cdots $ having $m_j$ elements of $S$ in the $j$th set. Now we can define the replica partition function:
\be
Z_n^{[S_1,S_2,\cdots]}=\prod_j \left(\sum_{\nu=1}^4 \prod_{k\in S_j}Z_n^{(\nu)}(k)\right)
\ee
Clearly the fermion spin structures are correlated within each subset $S_j$ but uncorrelated across different subsets. It is straightforward to verify that each of these is modular covariant, with a prefactor that can be eliminated as in \eref{zedenconj}.
A particular case is when each subset $S_j$ contains a single element $k_j$. In this case we find the uncorrelated replica partition function $Z_n^{\rm u}$ which we have argued is the correct one as $\ell\to 0$. Another particular case arises when there is a single subset, the entire set $S$. In this case one finds the correlated replica partition function $Z_n^{\rm c}$ which is correct as $\ell\to L$. All other cases lie somewhere in between. Therefore we conjecture that the full replica partition function of the free Dirac fermion is a linear combination (with $\ell$-dependent coefficients) over all partitions:
\be
Z_n=\sum_{\rm all~ partitions} a_{[S_1,S_2,\cdots]}(\ell/L)~ Z_n^{[S_1,S_2,\cdots]}
\ee

The physical intuition is that as the branch cut of length $\ell$ expands from 0 to $L$, the degree of correlation among replicas steadily increases. The coefficients $a$ must be such that only the terms corresponding to $Z_n^{\rm u}, Z_n^{\rm c}$ contribute at $\ell=0,L$ respectively. It will be interesting to pursue this direction and try to determine the coefficients. 

This discussion strongly suggests that the modular-invariant replica partition function for free fermions is a very complicated quantity (in striking contrast to the one for a single spin-structure, which is relatively simple and very explicit for all $\ell$). From this point of view it is encouraging that the free boson result of Ref.\cite{Chen:2015cna} is also very complicated! After all, the two are supposed to be equal on putting $R=1$ in the latter.

A possible application of the above ideas is to understand entanglement in more general conformal field theories, which at finite temperature and finite size remains largely an open problem. All the techniques used so far have been based on free field theory. In this context, it may be mentioned that minimal models and other CFT's do have free-field representations with a screening charge \cite{Dotsenko:1984ad} and their modular-invariant partition functions and correlators are quite well-known. It may be possible to determine replica partition functions for these theories, or at least their limiting behaviours for small and large intervals, using a combination of the free field representation and several physical principles including the relation to thermal entropy and the requirement of modular invariance.

\section*{Acknowledgements}

We would like to thank Tadashi Takayanagi for an extremely useful correspondence, and Shiraz Minwalla and Ashoke Sen for helpful discussions. This version has benefitted from helpful questions raised by the anonymous referee. The work of SFL is supported by an INSPIRE Scholarship, DST, Government of India, and that of SM by a J.C. Bose Fellowship, DST, Government of India. We thank the people of India for their generous support for the basic sciences.

\section*{Notation and conventions}

Our 2d theories are defined on a circular space of length $L$. The size of the interval over which we compute the \Renyi entropy is denoted $\ell$. The numbers $R_i$ describe the radii of the scalar field variables. We use conventions where T-duality is implemented by $R\to \frac{2}{R}$ and hence the self-dual radius is $R=\sqrt2$. The number of replicas is denoted $n$ and used to calculate the $n$th \Renyi entropy. The modular parameter is $q=e^{\pi i\tau}$ and the finite temperature case is described by $\tau=i\frac{\beta}{L}$ where $\beta$ is the inverse temperature. 

\bibliographystyle{JHEP}
\bibliography{modular}

\providecommand{\href}[2]{#2}\begingroup\raggedright\begin{thebibliography}{10}

\bibitem{Holzhey:1994we}
C.~Holzhey, F.~Larsen, and F.~Wilczek, {\it {Geometric and renormalized entropy
  in conformal field theory}},  {\em Nucl.Phys.} {\bf B424} (1994) 443--467,
  [\href{http://xxx.lanl.gov/abs/hep-th/9403108}{{\tt hep-th/9403108}}].

\bibitem{Casini:2004bw}
H.~Casini and M.~Huerta, {\it {A Finite entanglement entropy and the
  c-theorem}},  {\em Phys.Lett.} {\bf B600} (2004) 142--150,
  [\href{http://xxx.lanl.gov/abs/hep-th/0405111}{{\tt hep-th/0405111}}].

\bibitem{Calabrese:2004eu}
P.~Calabrese and J.~L. Cardy, {\it {Entanglement entropy and quantum field
  theory}},  {\em J.Stat.Mech.} {\bf 0406} (2004) P06002,
  [\href{http://xxx.lanl.gov/abs/hep-th/0405152}{{\tt hep-th/0405152}}].

\bibitem{2002Natur.416..608O}
A.~{Osterloh}, L.~{Amico}, G.~{Falci}, and R.~{Fazio}, {\it {Scaling of
  entanglement close to a quantum phase transition}},  {\em Nature} {\bf 416}
  (Apr., 2002) 608--610, [\href{http://xxx.lanl.gov/abs/quant-ph/0202029}{{\tt
  quant-ph/0202029}}].

\bibitem{Klebanov:2007ws}
I.~R. Klebanov, D.~Kutasov, and A.~Murugan, {\it {Entanglement as a probe of
  confinement}},  {\em Nucl.Phys.} {\bf B796} (2008) 274--293,
  [\href{http://xxx.lanl.gov/abs/0709.2140}{{\tt arXiv:0709.2140}}].

\bibitem{Riera:2006vj}
A.~Riera and J.~Latorre, {\it {Area law and vacuum reordering in harmonic
  networks}},  {\em Phys.Rev.} {\bf A74} (2006) 052326,
  [\href{http://xxx.lanl.gov/abs/quant-ph/0605112}{{\tt quant-ph/0605112}}].

\bibitem{Ryu:2006bv}
S.~Ryu and T.~Takayanagi, {\it {Holographic derivation of entanglement entropy
  from AdS/CFT}},  {\em Phys.Rev.Lett.} {\bf 96} (2006) 181602,
  [\href{http://xxx.lanl.gov/abs/hep-th/0603001}{{\tt hep-th/0603001}}].

\bibitem{Faulkner:2013ica}
T.~Faulkner, M.~Guica, T.~Hartman, R.~C. Myers, and M.~Van~Raamsdonk, {\it
  {Gravitation from Entanglement in Holographic CFTs}},  {\em JHEP} {\bf 1403}
  (2014) 051, [\href{http://xxx.lanl.gov/abs/1312.7856}{{\tt
  arXiv:1312.7856}}].

\bibitem{Casini:2005rm}
H.~Casini, C.~Fosco, and M.~Huerta, {\it {Entanglement and alpha entropies for
  a massive Dirac field in two dimensions}},  {\em J.Stat.Mech.} {\bf 0507}
  (2005) P07007, [\href{http://xxx.lanl.gov/abs/cond-mat/0505563}{{\tt
  cond-mat/0505563}}].

\bibitem{Cardy:2007mb}
J.~Cardy, O.~Castro-Alvaredo, and B.~Doyon, {\it {Form factors of branch-point
  twist fields in quantum integrable models and entanglement entropy}},  {\em
  J.Statist.Phys.} {\bf 130} (2008) 129--168,
  [\href{http://xxx.lanl.gov/abs/0706.3384}{{\tt arXiv:0706.3384}}].

\bibitem{Casini:2009sr}
H.~Casini and M.~Huerta, {\it {Entanglement entropy in free quantum field
  theory}},  {\em J.Phys.} {\bf A42} (2009) 504007,
  [\href{http://xxx.lanl.gov/abs/0905.2562}{{\tt arXiv:0905.2562}}].

\bibitem{Calabrese:2009qy}
P.~Calabrese and J.~Cardy, {\it {Entanglement entropy and conformal field
  theory}},  {\em J.Phys.} {\bf A42} (2009) 504005,
  [\href{http://xxx.lanl.gov/abs/0905.4013}{{\tt arXiv:0905.4013}}].

\bibitem{Casini:2008wt}
H.~Casini and M.~Huerta, {\it {Remarks on the entanglement entropy for
  disconnected regions}},  {\em JHEP} {\bf 0903} (2009) 048,
  [\href{http://xxx.lanl.gov/abs/0812.1773}{{\tt arXiv:0812.1773}}].

\bibitem{Calabrese:2009ez}
P.~Calabrese, J.~Cardy, and E.~Tonni, {\it {Entanglement entropy of two
  disjoint intervals in conformal field theory}},  {\em J.Stat.Mech.} {\bf
  0911} (2009) P11001, [\href{http://xxx.lanl.gov/abs/0905.2069}{{\tt
  arXiv:0905.2069}}].

\bibitem{Calabrese:2010he}
P.~Calabrese, J.~Cardy, and E.~Tonni, {\it {Entanglement entropy of two
  disjoint intervals in conformal field theory II}},  {\em J.Stat.Mech.} {\bf
  1101} (2011) P01021, [\href{http://xxx.lanl.gov/abs/1011.5482}{{\tt
  arXiv:1011.5482}}].

\bibitem{Headrick:2012fk}
M.~Headrick, A.~Lawrence, and M.~Roberts, {\it {Bose-Fermi duality and
  entanglement entropies}},  {\em J.Stat.Mech.} {\bf 1302} (2013) P02022,
  [\href{http://xxx.lanl.gov/abs/1209.2428}{{\tt arXiv:1209.2428}}].

\bibitem{Azeyanagi:2007bj}
T.~Azeyanagi, T.~Nishioka, and T.~Takayanagi, {\it {Near Extremal Black Hole
  Entropy as Entanglement Entropy via AdS(2)/CFT(1)}},  {\em Phys.Rev.} {\bf
  D77} (2008) 064005, [\href{http://xxx.lanl.gov/abs/0710.2956}{{\tt
  arXiv:0710.2956}}].

\bibitem{Herzog:2013py}
C.~P. Herzog and T.~Nishioka, {\it {Entanglement Entropy of a Massive Fermion
  on a Torus}},  {\em JHEP} {\bf 1303} (2013) 077,
  [\href{http://xxx.lanl.gov/abs/1301.0336}{{\tt arXiv:1301.0336}}].

\bibitem{Datta:2013hba}
S.~Datta and J.~R. David, {\it Renyi entropies of free bosons on the torus and
  holography},  {\em JHEP} {\bf 1404} (2014) 081,
  [\href{http://xxx.lanl.gov/abs/1311.1218}{{\tt arXiv:1311.1218}}].

\bibitem{Chen:2014hta}
B.~Chen and J.-q. Wu, {\it {Large Interval Limit of R\'enyi Entropy At High
  Temperature}},  \href{http://xxx.lanl.gov/abs/1412.0763}{{\tt
  arXiv:1412.0763}}.

\bibitem{Chen:2015cna}
B.~Chen and J.-q. Wu, {\it {R\'enyi Entropy of Free Compact Boson on Torus}},
  \href{http://xxx.lanl.gov/abs/1501.0037}{{\tt arXiv:1501.0037}}.

\bibitem{Cardy:2014jwa}
J.~Cardy and C.~P. Herzog, {\it {Universal Thermal Corrections to Single
  Interval Entanglement Entropy for Two Dimensional Conformal Field Theories}},
   {\em Phys.Rev.Lett.} {\bf 112} (2014), no.~17 171603,
  [\href{http://xxx.lanl.gov/abs/1403.0578}{{\tt arXiv:1403.0578}}].

\bibitem{Chen:2014unl}
B.~Chen and J.-q. Wu, {\it {Single interval Renyi entropy at low temperature}},
   {\em JHEP} {\bf 1408} (2014) 032,
  [\href{http://xxx.lanl.gov/abs/1405.6254}{{\tt arXiv:1405.6254}}].

\bibitem{Barrella:2013wja}
T.~Barrella, X.~Dong, S.~A. Hartnoll, and V.~L. Martin, {\it {Holographic
  entanglement beyond classical gravity}},  {\em JHEP} {\bf 1309} (2013) 109,
  [\href{http://xxx.lanl.gov/abs/1306.4682}{{\tt arXiv:1306.4682}}].

\bibitem{Faulkner:2013ana}
T.~Faulkner, A.~Lewkowycz, and J.~Maldacena, {\it {Quantum corrections to
  holographic entanglement entropy}},  {\em JHEP} {\bf 1311} (2013) 074,
  [\href{http://xxx.lanl.gov/abs/1307.2892}{{\tt arXiv:1307.2892}}].

\bibitem{Seiberg:1986by}
N.~Seiberg and E.~Witten, {\it {Spin Structures in String Theory}},  {\em
  Nucl.Phys.} {\bf B276} (1986) 272.

\bibitem{Cardy:1991kr}
J.~L. Cardy, {\it {Operator content and modular properties of higher
  dimensional conformal field theories}},  {\em Nucl.Phys.} {\bf B366} (1991)
  403--419.

\bibitem{Chen:2014ehg}
B.~Chen and J.-q. Wu, {\it {Universal relation between thermal entropy and
  entanglement entropy in CFT}},  \href{http://xxx.lanl.gov/abs/1412.0761}{{\tt
  arXiv:1412.0761}}.

\bibitem{Elitzur:1986ye}
S.~Elitzur, E.~Gross, E.~Rabinovici, and N.~Seiberg, {\it {Aspects of
  Bosonization in String Theory}},  {\em Nucl.Phys.} {\bf B283} (1987) 413.

\bibitem{DiFrancesco:1997nk}
P.~Di~Francesco, P.~Mathieu, and D.~Senechal, {\it {Conformal field theory}}, .

\bibitem{WolframTheta:}
functions.wolfram.com, {\it {Elliptic Theta}},  {\em
  {http://functions.wolfram.com/EllipticFunctions
  /EllipticTheta1/introductions/JacobiThetas/05/ShowAll.html}}.

\bibitem{Dotsenko:1984ad}
V.~Dotsenko and V.~Fateev, {\it {Four Point Correlation Functions and the
  Operator Algebra in the Two-Dimensional Conformal Invariant Theories with the
  Central Charge c \textless\ 1}},  {\em Nucl.Phys.} {\bf B251} (1985) 691.

\end{thebibliography}\endgroup

\end{document}